\documentstyle[aps,epsfig]{revtex}
\newcommand{\ccn}[2]{C^{\phantom{\dagger}}_{\mathstrut#1\mathbf{#2}}}    
\newcommand{\ccc}[2]{C^{\dagger}_{\mathstrut#1\mathbf{#2}}}              

\begin{document}
\title{New approach to the superconductivity problem}
\author{I.~M.~Yurin and V.~B.~Kalinin}
\address {Institute of Physical Chemistry, Leninskiy prosp. 31,
GSP-1, Moscow, 119991, Russia}
\maketitle
\begin{abstract}
 In this work, a question is tackled concerning the formation of a
superconducting condensate in an earlier  proposed model of "elastic jelly",
in which phonons of the ion system play the part of initiating ones.
It was shown that in distinction from the BCS theory, the momenta of forming
electron pairs are different from zero. This fact changes the pattern of the
description of the superconductivity phenomenon in the proposed model. First,
the gap in the one-electron spectrum appears due to the effect of a
"mean field" on the energy of electron state from the side of
occupied states, the nearest neighbors over the momenta grid. Second,
the condensate is formed by electron states with energies below
that of the gap edge, this is why the Fermi-condensation arises in
the system. Third, anomalous expectation values are strictly equal to zero
in the proposed model.
\newline
PACS number(s): 74.20.Mn, 74.72.-h 
\end{abstract}
\section{Introduction}
A considerable number of fundamental works 
have been devoted to tackling the superconductivity (SC) problem 
and, in  particular, a high temperature SC one (HTSC).
The opinion on the present state of the problem 
may be formed considering the works ~\cite{bib-1,bib-2,bib-3}.
Most relevant publication are to some extent associated 
with the BCS theory ~\cite{bib-4}. Now, for metals enumerated in
the Periodic System of elements (PSE), the SC theory
is considered to be more or less completed and the validity of the
BCS theory leaves no room for doubt on the part of the majority of
authors. At the same time in ~\cite{bib-5,bib-6,bib-7,bib-8} a conclusion
was drawn that the BCS theory may be not the only possible way to explain
the phenomenon of SC.

 Outstanding results ~\cite{bib-9} obtained by Bednortz J.G. and M\"uller K.A.
in 1986 for YBa$_{2}$Cu$_{3}$O$_{7-\delta}$ caused the great sensation among
chemists, physisists and material researchers. Since that time the highest
value for $T_s$ at about 164K was observed in 
HgCa$_{2}$Ba$_{2}$Cu$_{3}$O$_{8+\delta}$ ceramics. 

There are grounds available highlighting a possibility of observing high
$T_s$ in intercalates and nanotubes ~\cite{bib-3}. In our opinion, an
increase in the value of $T_s$ and the observation of HTSC are very
plausible in high-pressure (HP) phases. Almost twenty-five years ago,
SC was reported to have been observed in AuGa$_2$ ~\cite{bib-10}.
Comparatively high values of $T_s$ were observed in fullerenes,
HP phases with respect to graphite,
and their derivatives fullerides C$_{60}$M ~\cite{bib-11,bib-12}, where M stands
for K, Rb, Cs. The idea of obtaining HTSC in HP phases of Li$_{3}$P or
Li$_{3}$N is at present at the stage of being accepted ~\cite{bib-13,bib-14}.
In the last publications on this topic, SC is investigated in a HP phase of
MgB$_2$ ~\cite{bib-15,bib-16,bib-17}.
In conjunction with the problem under consideration, we also can't help
mentioning a hypothetical phase of metallic hydrogen ~\cite{bib-18} and
nitrogen (a private communication).

 The obtained results, especially the identification of HTSC with the
values of $T_{s}>100 K$, are rather inconsistent with both the predictions
of the BCS theory proper ~\cite{bib-19}, which had been made before
discovering HTSC, and any of its complimentary modifications, among
which the theory of bipolarons has to be singled out ~\cite{bib-2}.
In connection with this, we note that most HTSC materials
in stoichiometic phase are,
strictly speaking, dielectrics, whose conductivity is fully determined by
low concentrations of electroactive defects, whereas the theory
of bipolarons gives
low values of $T_s$ as the Fermi level approaches the edges of zones.
Thus, it does not explain HTSC for the aforementioned class
of compounds. 

In Russia, one usually considers the results of microscopic calculations
are in good agreement with the predictions of the BCS theory, including
the case of HTSC ~\cite{bib-20} as well. There is also another viewpoint
which is doubtful about the reliability of any fine calculation performed
by and large within the scope of the zone theory ~\cite{bib-21}. 

The problem is that a lot of poorly substantiated approximations did appear
for the years of the existence of the zone theory. Among these, for instance,
the procedure of accounting for the screening effects seems to be quite
harmless. These approximations were introduced by virtue of  quite relevant
reasons associated with an insufficient quick operation of computer
equipment to perform strict calculations of the zone structure of metals,
including that of the Periodic System. The application of these procedures,
however, renders the calculation of the electron-phonon interaction (EPI)
constants quite an unpredictable operation, depending only on theoretical
preferences of a specialist in the field of zone computations. 

From our viewpoint, the main source of difficulties consists in that all
the calculations associated with EPI pursue the goal of reducing the
Hamiltonian of the system to the Fr\"ohlich form. It has been known for
a long time ~\cite{bib-19} that the Fr\"ohlich model does not allow a
consistent accounting for the electron-electron interaction (EEI),
because the Coulomb interaction cannot still be reasonably separated
into that already accounted for in the initial values of Fr\"ohlich
Hamiltonian constants, and that which will thereafter show up in
the renormalization of the initial phonons owing to EPI.
Accordingly, all the microscopic  calculations of the EPI constants are,
strictly speaking, inconsistent.

Just these considerations underlied the need of a revision of the
fundamentals of the SC theory, which was undertaken in studies
~\cite{bib-22,bib-23,bib-24}. It was shown the attraction between
electrons exists in a long-wave range due to an exchange of virtual
phonons. This question was not discussed in earlier works (for
example, ~\cite{bib-25}) devoted to the problem being tackled.
We believe this was due to the following two basic causes.

First, in the framework of "procedure to account for the screening
effects" the function of the permittivity of electronic plasma was
incorrectly used for the removal of "undesirable" singularities
from the matrix elements of EPI in the long-wave limit. Second, the
simplified form of the effective EEI used in study ~\cite{bib-25}
did not suppose substantial differences in the electron interaction
both in the vicinity of the Fermi surface and far off it. 

 The authors of ~\cite{bib-25} well understood the disadvantages of the
aforementioned approaches. Remind that in ~\cite{bib-19} serious
forebodings were caused just by these approximations, which gave
rise to a detailed discussion. Moreover, an idea was put forward
that in a further development of the "jelly" model one should
have refrained from the use of ion-plasma oscillations as
initiating phonons.

 The accounting for just these remarks underlied the investigation
in the long-wave range of EEI, which had been undertaken in our
studies ~\cite{bib-22,bib-23,bib-24}, where the elastic oscillations of the ion
system serve as initiating phonons. It was shown the
adequate inclusion  of the phonon--phonon interaction in the
examination process enables one to avoid the application of poorly
substantiated procedures to "account for the screening effects".
And finally, the EEI potential was calculated without any
limitations as to its form.

It was shown that the unitary transformation effected in the framework
of the model suggested in ~\cite{bib-23,bib-24} can reduce the Hamiltonian
of the electronic system of an indefinite monoatomic metal to the following
form ($\hbar=1$):
\begin{eqnarray}
 \tilde{H}_{\mathrm tot}
 &=& \sum_{\mu} \int E_{\bf p}
     \ccc{\mu}{p}\ccn{\mu}{p}
      d{\mathbf p}
  + \tilde{H}_{ee}
  \label{eq-1} \\
 \tilde{H}_{ee}
 &=& \sum_{\mu}\sum_{\nu}\int\!\!\!\int\!\!\!\int 
     \tilde{U}({\bf p},{\bf k},{\bf q})
      \ccc{\mu}{p+q}\ccc{\nu}{k-q}
       \ccn{\nu}{k}\ccn{\mu}{p}
        d{\mathbf p} d{\mathbf k} d{\mathbf q}
 \label{eq-2}\ ,
\end{eqnarray}
and for $\tilde{U}({\bf p},{\bf k},{\bf q})$ at $p, k \approx K_F$ we have
\begin{equation}
 \tilde{U}({\bf p},{\bf k},{\bf q})
 =- 4 \left(\frac{zm}{3M}\right)^2
      \frac{G_{ee}}{q^2}
       \frac{K_F^2}{q^2-\chi_1}
        \frac{K_F^2}{q^2-\chi}
 \label{eq-3}
\end{equation}

where $E_{\bf p}=\frac{p^2}{2m}$,
$G_{ee}=\frac{e^2}{4\pi^2\epsilon}$,
$e$ is the electron charge,
$\ccc{\mu}{p}$ and $\ccn{\mu}{p}$
are operators of the creation and annihilation of electrons
with a momentum $\bf p$, respectively;
 $\mu$, $\nu$ are spin indices,
$\epsilon$ is the permittivity of the ion system,
 $M$ is the mass of ion,
 $m$ is the zone mass of electron,
 $z$ is the number of conductivity electrons per elementary cell;
$\chi=4\left(m^2\tilde{S}^2-\frac{zm}{3M}K_F^2\right)$,
$\chi_1=\left(1+\frac{zm}{3M}K_F^2/\lambda^2\right)\chi$,
$\lambda^2=(4e^2/\pi\epsilon)mK_F$,
$K_F$ is the value of Fermi vector, and $\tilde{S}$ is the observable
sound velocity in the system. 

The potential of the form ~(\ref{eq-3}) leads to an instability relative
to the formation of electron pairs. In distinction from the BCS theory,
the momenta of electron pairs are different from zero, while the pairs
bonding energy $E_b$ if a typical relationship $\frac{\chi}{2m} \ll E_b$
for SC systems is fulfilled, may be estimated as follows:
\begin{equation}
  E_b \sim \left(\frac{zm}{M}\right)^2\frac{G_{ee}K_F^4}{Q^3}
\label{eq-4}
\end{equation} 
where $Q=\chi^{1/2}$.

 It is obvious that $E_b\sim M^{-1/2}$, and this is consistent with the
observations of an isotopic effect in a series of metals ~\cite{bib-26}.
We also note that the coherence length $l_{coh}\sim Q^{-1}$.

\section{SC condensate in the model of elastic jelly}
The further discussion will be carried out for the final system - a cubic
crystal $L\times L\times L$ with periodic boundary conditions.
 
 Consider in the framework of Eqs.~(\ref{eq-1})-~(\ref{eq-3}) a wave
function of the bound state of an electron pair with maximum bonding
energy (MBE). This state will obviously be the state with zero spin.
With increasing $\tilde{S}$, its wave function will
first become hydrogen-like, and then delta-like, so that
at $\tilde{S} \to \infty$ it proves to be possible to derive for
the state $\left|{\bf R}\right\rangle$ of an electron pair with a momentum
${\bf R}$ at an accuracy of up to the normalizing factor the following
expression:
 \begin{equation}
\left|{\bf R}\right\rangle=\sum_{{\bf p}+{\bf k}=
{\bf R}}\tilde\delta({\bf k}-{\bf p})
\ccc{\uparrow}{p}\ccc{\downarrow}{k}\left|0\right\rangle
\label{eq-5}
\end{equation}
where $\tilde\delta({\bf k}-{\bf p})$ is a discrete delta-shaped function,
which is determined as follows: 
\begin{equation}
\tilde\delta({\bf r})=\left\{\begin{array}{l}
1, \quad \mbox{if}\quad|{\bf r_{\alpha}}|\le \frac{2\pi}{L}, \quad \alpha=x,y,z\\
0\mbox{ in the rest cases.}\\
\end{array}
\right.
\label{eq-6}
\end{equation}

Hence, it follows that in finite crystal, at least in the limit under
consideration $\tilde{S}\to\infty$, the following Hamiltonian ~(\ref{eq-7})
well describes the statistics and energy spectrum of MBE-states: 

\begin{equation}
H=\sum_{\mu, {\bf p}}
E_{p}\tilde{n}_{\mu {\bf p}}
-E_b\sum_{\bf p}\tilde{n}_{\uparrow\bf p}\tilde{n}_{\downarrow\bf p}
-\frac{E_b}{4}\sum_{\mu+\nu=0}
\sum_{\bf p\neq\bf k}\tilde\delta({\bf p}-{\bf k})
\tilde{n}_{\mu\bf p}\tilde{n}_{\nu\bf k}
+\frac{E_b}{4}
\sum_{\bf p\neq\bf k}\tilde\delta({\bf p}-{\bf k})
\tilde\ccc{\uparrow}{p}\tilde\ccc{\downarrow}{k}
\tilde\ccn{\uparrow}{k}\tilde\ccn{\downarrow}{p}
\label{eq-7}
\end{equation}
where $\tilde n_{\mu\bf p}=\tilde\ccc{\mu}{p}\tilde\ccn{\mu}{p}$
and the operators $\tilde\ccc{}{}$ and $\tilde\ccn{}{}$
are suggested to be linked with operators $\ccc{}{}$ and $\ccn{}{}$
from Eq.~(\ref{eq-1}) by the unitary transformation, whose derivation is
postponed for further publication.
Obviously the wave function of the ground state of the Hamiltonian
~(\ref{eq-7}) $\left|\Psi_0\right\rangle$ expressed in terms of
operators $\tilde\ccc{}{}$ does not differ from the analogous one in
the case of normal metal: $\left|\Psi_0\right\rangle=\prod_{k<K_F}
\tilde\ccc{\uparrow}{k}\tilde\ccc{\downarrow}{k}$.

The quasi-particle energy $\tilde{E}_{\mu\bf k}$ may be determined in
random phase approximation from the equation
$\left[H,\tilde\ccc{\mu}{k}\right]=\tilde{E}_{\mu\bf k}\tilde\ccc{\mu}{k}$
and thus one can get:
\begin{equation}
\tilde{E}_{\mu\bf k}=E_{\bf k}-2\tilde\rho_{\mu\bf k}\Delta_0
\label{eq-8}
\end{equation}
where $\Delta_0=7E_b$ and
\begin{equation}
\tilde\rho_{\mu\bf k}=\frac{1}{14}\left(\langle\tilde n_{-\mu\bf k}\rangle+
\frac{1}{2}\sum_{\bf p}\tilde\delta({\bf p}-{\bf k})
\langle\tilde n_{-\mu\bf p}\rangle\right)
\label{eq-9}
\end{equation}

For the large systems $(L\to\infty)$ in the absence of any spin order
in the electron system we have $\tilde\rho_{\mu\bf k}=
\langle\tilde n_{\mu\bf k}\rangle$. The averaged occupation number
$\tilde\rho_{\mu\bf k}$ may be expressed through the quasi-particle
energy $\tilde{E}_{\mu\bf k}$ by means of Fermi-Dirac distribution
function ($k_{B}=1$, $\xi$ is the chemical potential of electron system):
\begin{equation}
\tilde\rho_{\mu\bf k}=\left[1+
exp\left(\frac{\tilde{E}_{\mu\bf k}-\xi}{T}\right)\right]^{-1}
\label{eq-10}
\end{equation}
\begin{figure}[thb]
$$\psfig{figure=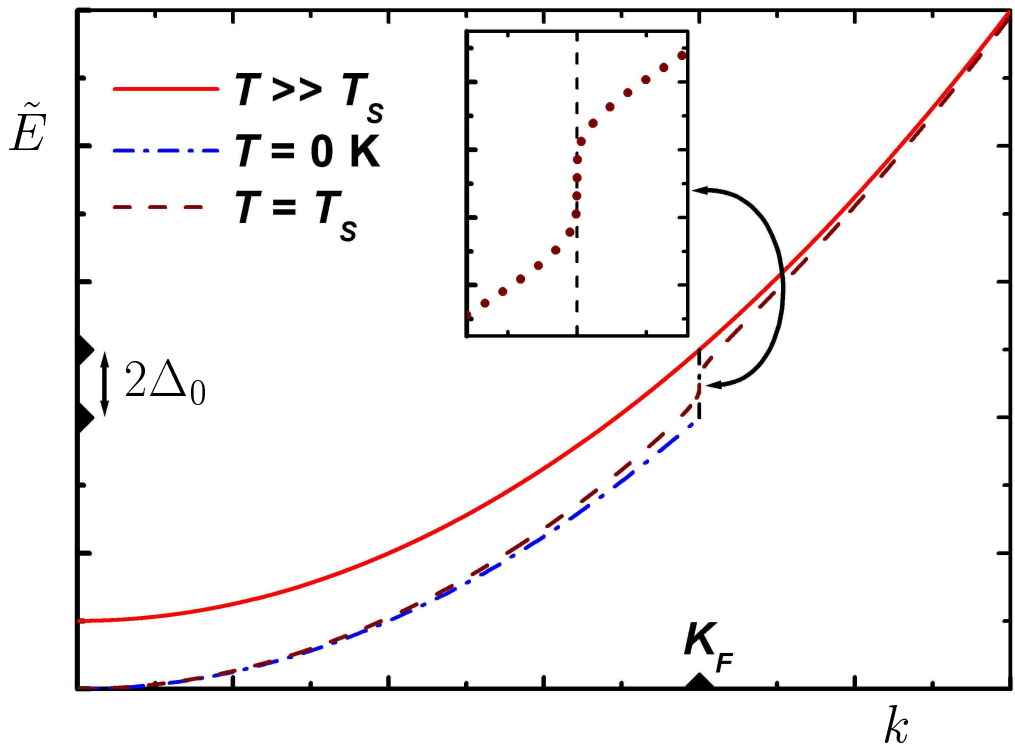,height=100mm}$$
   \caption{Temperature dependence of one-electron energy spectrum
in the model of elastic jelly (schematic)}  
\end{figure}
The evolution of a one-electron spectrum of the system with its temperature
varying becomes obvious from Fig. 1. At high temperatures, this spectrum is
described by an ordinary parabola; and at $T=T_s$ the spectrum exhibits a
peculiar feature which, at $T<T_s$, transforms into a gap. So, the SC
condensate is formed by electrons with the energy
$\tilde E_{\mu\bf k}<\xi$ at $T<T_s$.

Taking into account the Pauli principle, two versions of scatter are
possible in the system at lowest temperatures. In the first version,
the state of electron with momentum $k>K_F$ or hole with $k<K_F$ changes
(electron and hole concentrations in this case $\sim exp(-\Delta_0/T)$).
In the second version, an electron is removed from SC
condensate with the formation of an electron-hole pair,
or vice versa an electron-hole pair annihilates. The contributions of both
channels to collision integral are "frozen out" as temperature lowers
according to the exponential law. This circumstance combined with that
a shift of SC condensate as a whole in the momenta space leads to the
formation of a state with nonzero current is convincing in that a SC
transition is actually observed in the system under investigation.  

The transition temperature $T_s$ may be estimated by the equation
$\partial\tilde\rho/\partial{E}|_{\tilde{E}=\xi} = -\infty$.
Then we derive for $T_s$
\begin{equation}
T_s=\frac{\Delta_0}{2}
\label{eq-11}
\end{equation}

This estimation well agrees with the well known relationship
$T_s=2\Delta_0/3.5$, coming from the BCS theory in the case
of weak coupling. We suppose the deviations from the relationship
~(\ref{eq-11}) are due to correlation and many-particle effects. 

The remaining bound, but not MBE pair states may be accounted
for by inserting additional terms in the Hamiltonian~(\ref{eq-7}).
Yet this does not substantially distort the microscopic picture of the
phenomenon by virtue of two interrelated causes. Firstly, these terms
are small in their absolute values, compared with the bound energy of
MBE-state, which has already been accounted for. Secondly, the state of
the SC condensate is separated by an energy gap from the other states.
Therefore, in considering the state of SC condensate the arisen
terms may be accounted for in the framework of the perturbation theory
without any substantial changes in its wave function.

The behaviour of SC condensate in the electromagnetic field may be
described by the set of wave functions of MBE pair states due to the fact,
that the unitary transformation, linking operators $\tilde\ccc{}{}$ and
$\ccc{}{}$ involves these wave functions as transformation parameters.
This circumstance is helpful for the explanation of flux quantization
and Josephson effects in SC systems. The latter fact needs a detailed
consideration and is postponed for further publication. 

Finally, let consider the aforesaid limit $\tilde{S}\to\infty$. The energies of
the bound states of the starting Hamiltonian ~(\ref{eq-1}) are of an
analytical character in relation to the parameter $\tilde S$ throughout
the whole range of its admissible values. This is why the analyticity of the
energy parameters of the Hamiltonian ~(\ref{eq-7}) relative to the same
parameter $\tilde S$ cannot raise doubts. Then the presence of the
delta-shaped terms in the  Hamiltonian (\ref{eq-7}) for all the admissible
values of $\tilde{S}$ seems to be obvious.

\section{The condition for the onset of HTSC}
 A particular interest is spurred by the divergence of the bonding energy
$E_b$ at $Q \to 0$, following from Eq.~(\ref{eq-4}). In real systems,
this divergence should not reveal itself in the observed $T_s$ by virtue
of symmetry reasons  put forward in ~\cite{bib-23}. Simultaneously, with
varying parameters of specimens, there should be observed a maximum of the
transition temperatures if the following condition were fulfilled:
\begin{equation}
\tilde{S}^2 \approx \frac{zm}{3M}V_F^2
\label{eq-12}
\end{equation} 
This condition can be with a satisfactory accuracy fulfilled in systems
with variable physical parameters, in which, provided Eq. ~(\ref{eq-12}) is
obeyed, HTSC should be observed.
  
 Let consider a semiconductor with a rather low concentration of electroactive
defects, undergoing a phase transition with a change in volume. In both phases,
the material is stable, and, as a rule,
$\tilde S^2 \gg \frac{zm}{3M} V_F^2$. In unstable phases $\tilde S^2<0$.
It is possible to assume HTSC is observed in nonequilibrium systems, where
the phase transition is decelerated
$\tilde S^{2}\approx\frac{zm}{3M}V_F^{2}>0$
(the diamond-graphite transition is not realized exclusively due to
kinetic reasons).

One may expect the fulfilment of relationship~(\ref{eq-12}) with good
accuracy in the vicinity of the boundaries of stability of HP phases.
From this standpoint, it seems to be very promising to investigate the
kinetics of the phase transitions at HP with a view to searching for new
HTSC materials.

\section{Conclusions}
The basic qualitative differences of the BCS theory from the suggested
SC model are listed below.

1. In accordance with the BCS theory, the SC condensate is
formed by the pairs of electrons with zero momenta, and is therefore called
Bose condensate. 

In the proposed theory, the SC condensate is formed by electrons,
and, accordingly, transforms into the Fermi condensate.

2. In the BCS theory, the ground state of the system is obtained through
the superposition of the states with different total numbers of electrons.
It is this property that underlies one of the formulations of the BCS
theory ~\cite{bib-27} using anomalous expectation values
$\langle\psi^{\dagger}\psi^{\dagger}\rangle$, where $\psi^{\dagger}$
are electron creation operators, as order parameters.

Now, in the proposed theory these expectation values are strictly
equal to zero.
 
3. It is easy to show each of the states with a fixed total number of
electrons, forming the superposition of the ground state in the BCS theory,
is stationary state according to charge conservation law.
The energies of all these states are the same value $E_{min}$ due to the
fact, that the wave function of the ground state $\left|\Psi\right\rangle$
satisfies the following equation
$H\left|\Psi\right\rangle=E_{min}\left|\Psi\right\rangle$. Thus, a
conclusion may be drawn that the removal of electrons from the
superconductor does not change its energy.
Hence, the work function of electrons in SC at lowest temperatures
must be equal to zero.

Now, in the proposed model such a requirement is irrelevant.
\newline

Therefore, the investigation of the work function of electrons in SC materials
seems to be main practicable method for the experimental comparison of the
theories.

\section{Acknowledgements}
 We are grateful to Dr. N.V. Klassen, Dr. V.V. Gromov, and Prof. Allan Solomon
for their interest of our work and support of it.


\end{document}